\documentclass[sigconf,screen] {acmart}

\title{Affective Tools for Thought: Towards Shared Attention and Affective Reorienting in AI-Supported Thinking}

\author{Yifu Liu}
\affiliation{%
  \institution{UCL Interaction Centre}
  \city{London}
  \country{United Kingdom}
}
\email{yifu.liu.22@ucl.ac.uk}

\author{Raffaele Andrea Buono}
\affiliation{%
  \institution{UCL Interaction Centre}
  \city{London}
  \country{United Kingdom}
}
\email{raffaele.buono.18@ucl.ac.uk}

\author{Nadia Bianchi-Berthouze}
\affiliation{%
  \institution{UCL Interaction Centre}
  \city{London}
  \country{United Kingdom}
\orcid{0000-0001-8921-0044}
}
\email{n.berthouze@ucl.ac.uk}

\begin{abstract}
Current Tools for Thought (TfTs) treat affect as either friction that slows cognitive progress or a signal to optimise it. Drawing on enactive cognitive science, we argue that affect is constitutive of cognition: it reshapes the trajectory of thinking, not just the speed. We identify two core barriers for Affective TfTs---the lack of \emph{Shared Attention} (caring, directed attention to the user's mode of engagement) and the lack of \emph{Affective Reorienting} (the capacity to use emotional moments to open new trajectories rather than reinforcing predetermined ones)---and propose three design strategies that address both: Chain of Emotion $\times$ Chain of Thought, Affective Mirror, and Prompted Reorienting. The strategies are grounded in empirical findings from a study of a touch-aware conversational agent for embodied craft learning, and are oriented as provocations for future design.
\end{abstract}


\begin{document}

\maketitle

\section{Introduction}


Current Tools for Thought (TfT)---AI systems designed to augment human cognition rather than merely automate tasks \cite{tankelevitch2025synthesis}---operate on an implicit assumption: that thinking proceeds along a roughly linear trajectory toward a goal. Within this framework, AI functions as an instrumental support, a tool to aid the user along \textit{the} trajectory more effectively. The destination, and oftentimes the path itself, is a given. Within such framework (\textit{instrumental progress model}), affect is either understood as friction and noise that slows down progress, or an informative signal to leverage to improve adherence. 

This paper challenges this model, and the role of affect within it, from an enactivist standpoint \cite{hila2026enactivist}. Affect does not merely speed up, or slow down, thinking along a fixed path. We argue that affect, as an enactive scaffold to cognition, has the potential to reshape paths and user journeys, beyond instrumentalist impositions which frame tasks as ready-made objectives. For instance, a frustrated learner is not just experiencing a cognitive burden to be smoothened over; similarly, an eager learner is not merely moving along a path quicker, with tools being designed to support such increase in speed. Rather, we argue that affect foregrounds affordances and potentialities \cite{nygren2024embodied} which radically reshape the activity space itself, and allow tasks to move along new gradients. We argue that by reframing AI as Affective Tools for Thought could support \textit{divergent trajectories} for thinking. These emerge when affect is taken as a scaffold to cognition, reframing progress away from the execution of a single, predetermined outcome and loop. 

We make two contributions. First, we articulate a theoretical grounding for why affect is constitutive of, not incidental to, thinking---and what this means for TfT design (\S\ref{sec:theory}). Second, we identify two core barriers for Affective TfTs (\emph{Shared Attention} and \emph{Affective Reorienting}), and propose three design strategies that address both, grounded in empirical findings from the study of a touch-aware conversational agent \cite{selfcite} (\S\ref{sec:framework}). The strategies are provocations for future design, with their limitations discussed in (\S\ref{sec:discussion}). 




\section{Theoretical Grounding: Affect and Thinking}
\label{sec:theory}

\begin{figure*}[t]
  \centering
  \includegraphics[width=\linewidth]{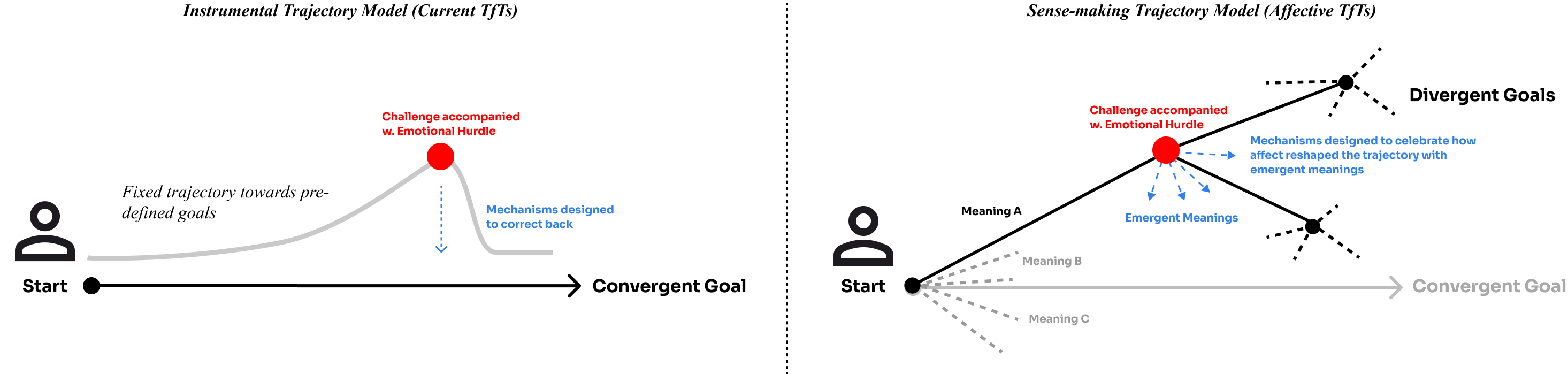}
  \caption{(Left) The Instrumental Trajectory Model common in current TfTs, where affect is treated as a hurdle to be corrected to reach a predefined goal. (Right) The proposed Sense-making Trajectory Model, where affective hurdles act as scaffolds that reshape thinking toward emergent, divergent goals.}
  \label{fig:mapping}
\end{figure*}

Research on Tools for Thought has drawn on ideas of reflective thinking to ground the design of AI systems that augment cognition \cite{singh2023mind, tankelevitch2025synthesis}. While informative, the affective implications behind these conceptualisations have been often set aside. For instance, Singh et al. \cite{singh2023mind} cite Dewey's understanding of perplexity as a prerequisite for critical thinking. Perplexity is however treated as a purely cognitive state, rather than as an embodied and extended affective experience that may trigger and sustain reflective enquiry. The TfT synthesis provided by Tankelevitch et al. \cite{tankelevitch2025synthesis} acknowledges this gap, noting that, while most current work augments deliberate, high-level, 'System 2' reasoning, leveraging emotions and other 'System 1' processes remains an underdeveloped frontier. We argue however that such dichotomised framing of cognition and affect, System 2 and System 1, presents the two as ultimately separate channels intersecting \textit{a posteriori}.

This assumption has shaped how existing TfT systems are designed. Zhang and Reicherts' \emph{ExtendAI} \cite{zhang2025extendai} asks users to externalise their reasoning before receiving AI feedback---a form of `process-oriented support' that reads the user's thinking rather than delivering end-to-end solutions. Cheung's \cite{cheung2025ignorant} `ignorant co-learner' deliberately withholds knowledge to create cognitive friction, fostering moments of uncertainty, dissonance, or pause that compel users to think critically. Gmeiner et al.'s \cite{gmeiner2023exploring} metacognitive support agents proactively ask reflective questions during design tasks to leverage affect. And while these all understand cognition as something AI can directly engage with, as a Tool for Thought, the studies presented all understand reasoning and cognition as the main objects of support: affect is understood as merely an external parameter, either as noise to be filtered out, or a signal to adjust to to aid the user in task completion.

\paragraph{Three positions on affect and cognition.} We see the approaches to TfT as responding to two main ways of positioning affect. Against these, we propose a third position.

\textbf{Position~1---Affect as friction.} The Cartesian view which frames affect and bodies as pitted against rational thinking and cognition. Here, TfT should optimise information processing, and minimise emotional disruption. 

\textbf{Position~2---Affect as instrument.} This view, epitomised by most affect automatic recognition systems, treats emotions as detectable and tractable signals that can be leveraged to optimise performance by building semi-adaptable systems \cite{damasio1994descartes, picard1997affective}. A system here detects frustration and reduces task difficulty; it designs \textit{around} affect, rather than \textit{through} it.

We see both positions as \textit{instrumentalising} affect, as subservient to cognition. We thus argue that both positions frame a system of individual, self-contained cognizers, with AI functioning as an optimising tool for adherence to predefined paths of reasoning. Cognitive reasoning follows linear, individual trajectories which are either hampered, or supported, by affect as an externality.

\textbf{Position~3---Affect as constitutive scaffold.} We argue that affective modes of engagement are constitutive of thinking, thus opening up ways to rethink tasks and goals. We argue that affect is not a variable within thinking, or an externality to plans: rather, it is the very embodied and situated resource through which worlds are brought about and provisionally constituted. In this sense, we draw on the enactivist idea of sense-making, as an agent's embodied and affectively-charged activity of bringing forth a world of significance through their ongoing coupling with an environment and other agents \cite{varela1991embodied, dipaolo2010horizons, dipaolo2012enactive}. Affect scaffolds embodiment and cognition \cite{gallagher2017}: it produces attentional orientations, interpretive possibilities, significative potentialities, multiple relationships to a world and the tasks emergent from it. Affect thus does not simply \textit{influence} how a task is performed, but actively \textit{shapes} what can count as a task, what can be perceived as relevant, and what kinds of actions become intelligible or viable in a given situation. It is through affective attunement that certain features of the environment come to matter at all, and that particular problem framings, goals, and criteria of success take form.

\paragraph{From \textit{instrumental progress} to \textit{sense-making}.} 

We then propose a rethinking of TfT away from the ideal of instrumental progress, and instead fostering affect-based multiplicity. We do this by arguing that affect is not merely a factor `in' cognition: instead, it is a central engine to cognising, as the capacity to construct worlds through sense-making. Here, agents (humans and AI itself) and the world are not pre-given entities externally interacting, but they are rather mutually constituted through engagements \cite{thompson2007mind} which intrinsically link them, based on forms of shared attention and care---i.e., intersubjectivity \cite{mcgann2014, hutto2017evolving}. 
From this premise, we conclude that tasks themselves, their trajectories and benchmarks for success, are emergent through affect and affect-driven sense-making. When a learner's frustration transforms the meaning of a task from ``an interesting challenge'' into ``evidence of my inadequacy,'' this opens up a space to precisely support the user where the task could lead them, rethinking the entire landscape of what is cognitively possible. In our empirical study \cite{selfcite}, we observed this directly: chatbot-led reflections did not simply restore participants to their original task trajectory. Rather, affect re-pivoted thoughts. During guided reflection, one participant arrived at the experience of repair as ``almost meditative'' (P12); another, prompted to reconsider tactile assumptions, discovering unexpected relationships with the material (P15); a third, when helped by the AI to notice they were rushing, began to reconceive the activity as something to attend to, rather than merely complete (P6). In each case, affect-aware interaction opened a new trajectory, rather than correcting a deviation. 
This reframes the role of a TfT from an instrument that optimises progress and productivity along predefined trajectories, to a participant in sense-making that can open, rather than close down, alternative ways of understanding the task, and oneself within it.

From these premises, we start drawing out some preliminary design stakes and implications: (1) TfTs that assume affect to be a constitutive scaffold should treat emotional moments---frustration, confusion, surprise, doubt---not as obstacles to overcome, but as potential pivot points where the meaning of the task might productively shift, and with it, perhaps, the task itself and the world within which the task comes to be configured; (2) TfTs as relational systems that support shared attention and concern, treating interaction not as the exchange of inputs and outputs but as a process through which both the user and the system can reshape how the task is understood. In this view, meaningful support arises from sustaining intersubjective engagement---moments of noticing, pausing, and reorienting together---through which divergent trajectories of thinking can become possible.

\section{Framework: Affective Tools for Thought}
\label{sec:framework}

The framework has two components. First, we identify two core barriers that current TfTs do not address: \emph{Shared Attention} and \emph{Affective Reorienting} (\S\ref{sec:problems}). Second, we propose three design strategies, Chain of Emotion $\times$ Chain of Thought, Affective Mirror, and Prompted Reorienting , each of which addresses both barriers differently (\S\ref{sec:strategies}, Figure~\ref{fig:mapping}). We discuss their limitations in \S\ref{sec:discussion}.

Our framework builds on the theoretical foundations drawn in (\S\ref{sec:theory}) by specifically designing for ``care'', and against ``Loop-in-the-Human''. 
The first idea comes from Heidegger's concept of \textit{Sorge} \cite{elleybrown2021}: here, attention is not merely a form of detection, monitoring and response, as current affect recognition systems provide. Rather, care is a form of intersubjective orientation: carer and cared for interlock each other, and transform each other towards the constitution of a common trajectory. For instance, a human tutor who notices a student struggling does not merely register the struggle and adjust parameters; they attend with concern, drawing on their own experience to pose questions that are personally authentic and oriented toward who this person is becoming. This caring attention is constitutive of a kind of support that enables collective (re)making, and is precisely what current AI systems lack.

We borrow the second idea from the work of Gambarotto and colleagues to argue that current Human-in-the-Loop strategies reduce human purposiveness to quantifiable costs and benefits. The result is that the ``Human-in-the-Loop increasingly risks becoming a Loop-in-the-Human'' \cite{gambarotto2025design}: a configuration where human thought is shaped to resemble the algorithmic systems it was meant to oversee. In TfTs, this means systems designed to support thinking may constrain it, channelling users along predetermined trajectories and treating deviations as errors.

\subsection{Affective TfT Barriers}
\label{sec:problems}
\begin{figure*}[t]
  \centering
  \includegraphics[width=\textwidth]{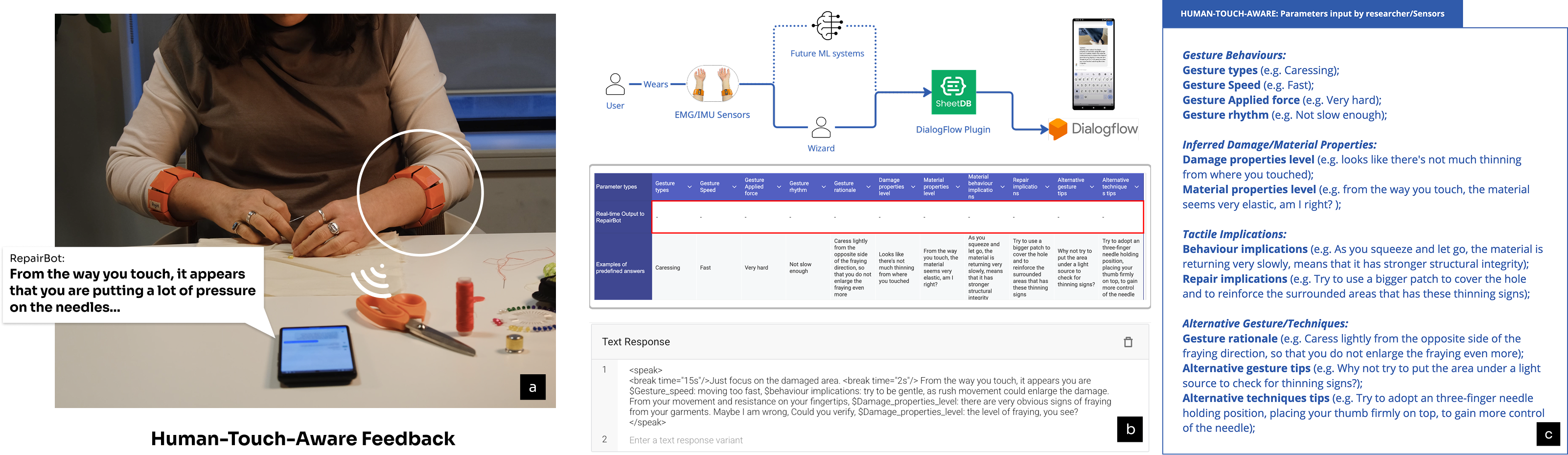}
  \caption{The RepairBot Feedback System; an overview of touch-aware interaction where the system analyzes gesture behaviors to provide inferred material properties and alternative techniques as affective mirrors.}
  \label{fig:mapping}
\end{figure*}

The identified barriers builds on the scenarios and empirical findings from our study of human-AI interaction in embodied craft learning \cite{selfcite}, a Wizard-of-Oz system using a touch-aware chatbot to support participants repairing garments.

\textbf{Barriers~1: The Lack of Shared Attention.} By \emph{Shared Attention} we mean not merely that the system monitors the user, but that it attends to the user's \emph{affective mode of engagement}---the way they are relating to the task emotionally in the moment. This involves three components: (i) \emph{noticing} \textbf{how} the person is engaging, not just \textbf{what} they are doing; (ii) \emph{interpreting} that engagement within the developing trajectory---reading the present moment in light of what has come before; and (iii) \emph{making that attention felt}---communicating, through the quality of its responses, that the person's mode of engagement matters.

Instead of capturing or sensing the entire context, Participants described a human tutor as someone who not only ``sense the critical points'' and offer proactive support, but ``makes the action feel seen''(P4)\cite{selfcite}. This names the quality of \emph{Sorge}: a directed, caring attention that differs from tracking or monitoring. Current TfTs provide sophisticated cognitive support, but not this quality of \textbf{shared} attention: for instance, current TfTs \emph{track behaviour}, such as keystrokes, reasoning patterns, task completion \cite{zhang2025extendai} \cite{cheung2025ignorant} \cite{gmeiner2023exploring}---without attending to how the person \emph{relates} to what they are doing. Without Shared Attention, the system operates on a model of cognition that is disconnected from the person's lived experience: it sees the task but not the person doing it. 

\textbf{Barriers~2: The Lack of Affective Reorienting.} By \emph{Affective Reorienting} we mean the capacity to use moments of emotional intensity—frustration, resistance, self-doubt, but also unexpected delight or curiosity—as pivot points where the person's relationship to the task might shift. The key distinction is between loop-reinforcing responses, which treat affect as incidental and adjust parameters within the existing trajectory (e.g., reducing difficulty when frustration is detected), and reorienting responses, which attend to what the emotional moment reveals about how the person relates to the task. Current TfTs lack this second capacity.

In our study \cite{selfcite}, participants experienced self-reinforcing cascades where frustration produced self-doubt, self-doubt produced rushing, and rushing produced more errors—a spiral that led some to disengage entirely, reaching what they described as "saturation points" or declaring themselves "not gifted at this." This was not merely an emotional event, but a shift in what the task meant for the person: from "can I learn this?" to "am I the kind of person who can do this?". Our chatbot responses remained loop-reinforcing: when P7 expressed resistance, we provided easier tasks—adjusting difficulty but preserving the same evaluative frame. A \textbf{reorienting response} would have treated resistance not as an obstacle but as an opportunity, inviting a reconsideration of what the difficulty signified, and what it had opened up.

More broadly, current proactive support mechanisms \cite{liu2025proactive} demonstrate the value of anticipatory assistance, but operate within the existing trajectory, smoothing the path before difficulty arrives. They make the loop more efficient; they do not question whether the loop is the right one. What is missing is the capacity to recognise when emotional intensity signals that the trajectory itself needs to change—and to stay with the person at that moment rather than resolving it. Without this capacity, the system risks becoming what Gambarotto et al.\ call a Loop-in-the-Human \cite{gambarotto2025design}.

\begin{figure}[t] 
  \centering
  \includegraphics[width=\columnwidth]{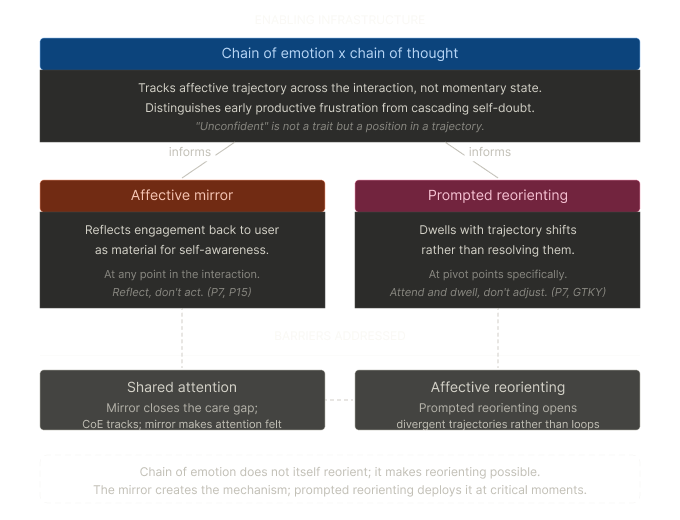} 
  \caption{The Affective TfT Framework; mapping the two core identified barriers to three proposed design strategies.}
  \label{fig:fig_affective_tft_framework_diagram}
\end{figure}

\subsection{Affective TfT Strategies}
\label{sec:strategies}

\begin{figure}[t] 
  \centering
  \includegraphics[width=\columnwidth]{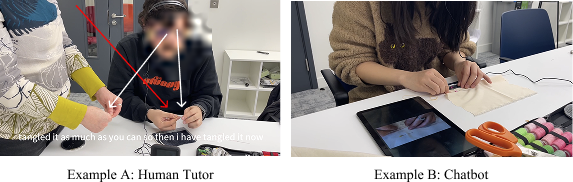} 
  \caption{Shared Attention in Context; comparing the intersubjective orientation of a Human Tutor (A) and the Chatbot (B).}
  \label{fig:mirror}
\end{figure}

\begin{figure*}[t]
  \centering
  \includegraphics[width=\textwidth]{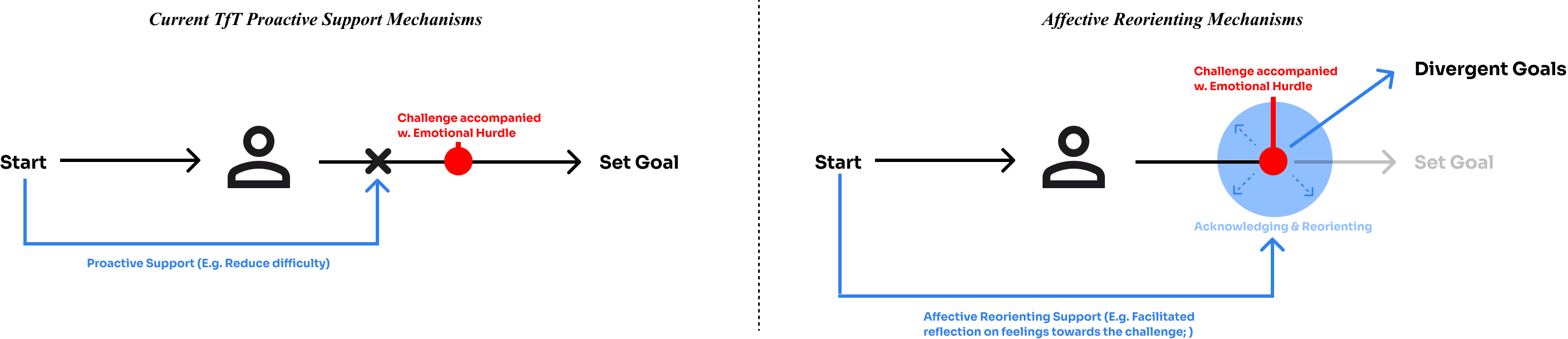}
  \caption{(Left) Current proactive mechanisms that reduce difficulty to keep users in a "loop". (Right) Prompted Reorienting mechanisms that dwell with the person at pivot points, allowing the meaning of the task to shift productively.}
  \label{fig:mapping}
\end{figure*}

\paragraph{\textbf{Affective Mirror.}} The Affective Mirror directs the system's reading of the user's engagement back to the user, creating occasions for affective self-awareness. Rather than the system acting on what it senses, it makes the user's own mode of engagement visible to them, not as a diagnostic label, but as material for reflection.

We arrive at this strategy through empirical observation. In our study, moments where the system reflected aspects of engagement back to participants, even imperfectly, generated unexpected self-awareness. P7 found incorrect chatbot feedback useful because it prompted self-examination. When our touch-aware system offered an inference about P15's material interaction, the encounter led P15 to re-examine her own tactile perception and discover an unexpected relationship with the material. This happened not because the system was right, but because the reflection opened a new way of attending. Through facilitated noticing, P12 arrived at the experience of repair as ``a very calm, almost meditative experience,'' and P6 came to see they were ``rushing ahead instead of enjoying the process.'' None of these outcomes were pre-designed; they emerged through the encounter between system and person.

These observations align with existing work: research on body movement and touch has demonstrated that embodied behaviour expresses not merely task actions but affective engagement; the body's involvement shapes experience itself \cite{bianchiberthouze2016}. Yet users are often unaware of what their engagement reveals. H\"{o}\"{o}k's \emph{soma design} \cite{hook2018designing} argues that technology can cultivate this bodily self-awareness, and Sanches et al.\ find that \emph{ambiguous} bio-feedback prompts more reflection than precise diagnostic labels \cite{sanches2019ambiguity}. For future TfTs, the Affective Mirror implies a specific design orientation: when the system senses a shift in engagement, its first response should be to \emph{reflect} rather than \emph{act}, surfacing what it has noticed in open-ended, non-diagnostic language (``I notice you've slowed down and are handling the material differently'') rather than classifying an emotion or adjusting a parameter. This addresses Shared Attention by closing the care gap: if the system cannot attend with genuine concern, it can help the \emph{user attend to themselves}. It also creates the conditions for Affective Reorienting. As P12's and P6's shifts suggest, the act of seeing one's own engagement can itself become the moment where the meaning of the task changes.

\paragraph{\textbf{Chain of Emotion $\times$ Chain of Thought.}} Chain of Emotion proposes that the system's reasoning process (its Chain of Thought \cite{wei2022chain}) must be informed by a temporally-structured model of the user's affective engagement, a Chain of Emotion, where each transition is shaped by the preceding one. Rather than classifying a momentary emotional state, the system tracks how the person's mode of engagement has been evolving, and lets that trajectory inform what kind of response is appropriate.

In our study, the participants who disengaged did not simply ``become frustrated.'' They moved through a trajectory: curiosity gave way to confusion, confusion to frustration, frustration to self-doubt, and self-doubt to withdrawal. This aligns with D'Mello and Graesser's model identifying confusion-to-frustration transitions as critical junctures in learning \cite{dmello2012dynamics}. Crucially, the same emotion, frustration, called for different responses depending on where it sat in the trajectory. Early frustration during exploration signalled productive challenge; frustration following repeated failure and self-doubt signalled a cascade toward disengagement. Without tracking the trajectory, the system cannot distinguish between the two. Prather and Reeves observe that ``underprepared, unconfident, and underperforming students seem to benefit the least'' from AI tools \cite{prather2025underprepared}. Chain of Emotion reframes this: \emph{``unconfident''} is not a trait but a position in a trajectory.

For future TfTs, this strategy implies that the system's reasoning architecture must maintain an evolving model of affective engagement across the interaction, not a snapshot of the current state. The question shifts from ``the user is frustrated, reduce difficulty'' (Position~2) to ``what role is this person's frustration playing in their thinking, and what kind of attention does it call for?'' For Shared Attention, this is the mechanism by which the system approximates \emph{attuning}: reading the present moment within the developing trajectory. For Affective Reorienting, it is the enabling infrastructure. Without it, the system cannot recognise when the emotional trajectory signals that the loop itself needs to change. Chain of Emotion does not itself reorient; it makes reorienting possible.

\smallskip

\paragraph{\textbf{Prompted Reorienting.}} Where the Affective Mirror reflects engagement at any point in the interaction, Prompted Reorienting acts at a specific juncture: when the Chain of Emotion signals a trajectory shift, the system responds not by adjusting parameters but by \emph{dwelling with} the shift itself. The distinction is between resolving an emotional moment (reducing difficulty, offering encouragement) and treating it as a pivot point where the person's relationship to the task might change. This contrasts with both reactive and proactive support \cite{liu2025proactive, gmeiner2023exploring}, which operate within the existing trajectory. Both make the loop more efficient; neither questions whether the loop is the right one.

We arrive at this strategy based on empirical findings. When P7 expressed resistance, we scaffolded the difficulty, a loop-reinforcing response that preserved the evaluative frame. A reorienting response would have stayed with the resistance: \emph{what does this difficulty mean to you? Could it be approached as exploration rather than test?} Conversely, the ``Getting to Know You'' module asked participants to reflect on a garment's emotional history before technical work. We designed the prompt, but not what followed: P12 arrived at repair as ``a very calm, almost meditative experience,'' and P6 reconceived their relationship to the activity entirely. What these moments shared was that something shifted in the person's engagement, and the interaction did not smooth past it.

For future TfTs, Prompted Reorienting implies a design rule: at moments where the Chain of Emotion registers a trajectory shift, the system's default should be to \emph{attend and dwell} rather than \emph{adjust and resolve}. In practice, this means posing questions that make the shift itself visible (``You were experimenting freely a moment ago, something seems to have changed'') rather than acting on the inferred emotion. Ayyappan and Joyner's mode-switching \cite{ayyappan2025selfreliance}, which lets students toggle between ``tell me the answer'' and ``ask me a question,'' gestures toward giving the user agency at such moments, but assumes the task's meaning is stable. Prompted Reorienting recognises that the meaning itself might shift, and that the system's role is not to engineer that shift, but to refuse to foreclose it.

\section{Discussion}
\label{sec:discussion}

\paragraph{How the two barriers interact.} Without Shared Attention, without the system attending to the person's affective engagement, the system cannot recognise when Affective Reorienting is called for. And without Reorienting, even the most attentive system will reinforce the loop: reading engagement accurately but responding only within the existing trajectory. The two barriers need each other's solutions. Chain of Emotion provides the temporal model that makes Shared Attention possible; the Affective Mirror creates the reflective mechanism through which attention becomes felt; and Prompted Reorienting deploys that mechanism at the moments where it matters most. None functions without the others.

\paragraph{Divergent pathways in constrained scenarios.} A question of scope deserves direct address: how should an Affective TfT behave in fixed-goal domains, such as mathematics tutoring, where success criteria are stable, and ``divergence'' risks becoming distraction? Our answer is that the divergence we propose does not operate at the level of the task's \emph{epistemic content}: the correct answer remains correct. It operates at the level of the learner's \emph{relationship} to the task, and that relationship is never fully determined by the task's success criteria alone. In practice, a fixed-goal Affective TfT would not question whether a student should be learning multiplication. It would instead recognise, through Chain of Emotion, when frustration has shifted from ``I don't understand this yet'' to ``I am not capable of understanding this.'' A TfT tool could open a different trajectory: naming that shift, reflecting it back through the Affective Mirror, and reorienting the student's relationship to difficulty itself. Two students who both eventually reach the correct answer are not, in this sense, on equivalent paths. The degree of reorientation a system enables remains a design decision, appropriately constrained in fixed-goal contexts; but the underlying barriers, Shared Attention and Affective Reorienting, remain live across both.

\paragraph{Rethinking evaluation.} Standardised evaluation assumes that success criteria can be specified in advance. Affective TfTs are designed for situations where the learner's relationship to those criteria is itself what is at stake, unsettling linear evaluation pathways. Merely measuring task performance, goal completion or adherence would miss the point: it would treat as settled exactly what the framework holds open. In lieu of linear measurable outcomes, we argue that what may come under scrutiny is whether a productive, generative reorienting has occurred. For instance, trajectory divergence (whether a person's framing of the task shifts across an interaction) offers a direct index of this, assessable through pre/post interviews or discourse analysis. Affective transition patterns could track whether interventions interrupt negative cascades without resolving them into task completion, a meaningful signal precisely because it decouples affective shift from performance outcome. Self-reported relationship to difficulty and inter-rater reliability on identified pivot moments in transcripts are both methodologically credible moves in the same direction. What these measures share is that they index change \emph{in orientation}, rather than arrival at a destination. This shift in evaluative logic matters, because it holds evaluation itself within the framework the paper proposes: an Affective TfT that is only legible through task performance has already conceded the ground it set out to contest.

\paragraph{The boundary with affective computing.} Our framework draws on sensing and inference mechanisms developed within the affective computing tradition; what changes is not the sensing but what the system does with what it senses. Classic affective computing detects an emotion and responds to it \cite{picard1997affective}. H\"{o}\"{o}k's affective loop framework \cite{hook2009affective} goes further, treating emotion as co-constructed in interaction and keeping interpretive agency with the user. Our framework extends this lineage by arguing that affect does not merely shape the interaction but can reshape what the task \emph{is}. The practical boundary is less clean than the ontological one: an implemented Affective TfT would still classify engagement patterns and make inferences. The difference lies in what follows. Where an adaptive system uses inference to adjust parameters within a stable frame, an Affective TfT uses inference to recognise when the frame itself may need to shift, and then stays with that moment rather than resolving it.

\paragraph{The authenticity gap.} A human tutor draws on personal experience to pose questions with individual authenticity. Chain of Emotion may restructure the system's computational priorities to emulate the \emph{structure} of concern, attending to what matters to the person, tracking what is at stake. But the question remains whether restructuring priorities is the same as genuine care. Participants valued the chatbot's non-judgmental quality \cite{cheung2025ignorant, pilcher2025productive}, a genuine affordance. But we must acknowledge what is lost: the authentic collision of different lived perspectives, the way a tutor's genuine surprise opens a space no algorithm can replicate. This tension, between AI's non-judgmental consistency and the irreplaceable generativity of authentic encounter, is not a problem to solve but a design space to navigate.

\paragraph{Limitations.} Our empirical grounding comes from a single study in embodied craft learning with a Wizard-of-Oz system: the attuning was performed by a human researcher, not an automated system. Transferability to text-based TfTs is theoretically motivated but not demonstrated. The Affective Mirror cannot substitute for encountering a genuinely different perspective. Prompted Reorienting is the most speculative strategy: we observed pivots emerging incidentally and acknowledge that designing systems to stay with emergent moments, without becoming prescriptive, is a significant challenge. The design principles we propose are derived from retrospective analysis of our empirical data; they have not yet been implemented or evaluated in an automated system. Cross-session Chain of Emotion modelling raises questions about data persistence, privacy, and reductive emotional profiling.





\section{Conclusion}
We have identified two core barriers for Affective Tools for Thought: the lack of Shared Attention, caring, directed attention to the user's mode of engagement, and the lack of Affective Reorienting, the capacity to use emotional moments to open new trajectories rather than reinforcing predetermined ones. We proposed three design strategies: Chain of Emotion as enabling infrastructure that tracks affective trajectory across interactions; the Affective Mirror, which reflects engagement back to users as material for self-awareness; and Prompted Reorienting, which dwells with trajectory shifts rather than resolving them. The strategies are grounded in empirical findings from a touch-aware conversational agent for embodied craft learning, and are oriented as provocations for future design rather than implemented solutions. We invite TfT researchers to consider: what would our tools look like if they were designed not to keep users on track, but to stay with them when something unexpected emerges?

\bibliographystyle{ACM-Reference-Format}

\end{document}